\documentclass[12pt]{article}
\usepackage[utf8]{inputenc}
\usepackage[
    backend=biber,
    maxnames=4,
    style=ieee,
]{biblatex}
\pdfoutput=1

\usepackage{amsmath}
\usepackage{amssymb}
\usepackage{units}
\usepackage{graphicx}
\usepackage{float}
\usepackage{slashed}
\usepackage{array}
\usepackage[leftcaption,raggedright]{sidecap}
\usepackage{caption}
\usepackage[hidelinks]{hyperref}

\newcommand\pubnumber{NuPhys2026-Ellie-O'Brien}
\newcommand\pubdate{\today}

\def\sheff{School of Mathematical and Physical Sciences \\ University of Sheffield, Sheffield S3 7RH, United Kingdom}

\def\support{\footnote{
  On behalf of the Hyper-Kamiokande Collaboration.
}}

\def\Title#1{\begin{center} {\Large #1 } \end{center}}
\def\Author#1{\begin{center}{ \sc #1} \end{center}}
\def\Address#1{\begin{center}{ \it #1} \end{center}}

\newcommand\pubblock{\rightline{\begin{tabular}{l} \pubnumber\\
         \pubdate  \end{tabular}}}
\newenvironment{Abstract}{\begin{quotation}  }{\end{quotation}}
\newenvironment{Presented}{\begin{quotation} \begin{center} 
             PRESENTED AT\end{center}\bigskip 
      \begin{center}\begin{large}}{\end{large}\end{center} \end{quotation}}
\def\Acknowledgements{\bigskip \bigskip \begin{center} \begin{large}
             \bf ACKNOWLEDGEMENTS \end{large}\end{center}}

\def\beq{\begin{equation}}
\def\eeq#1{\label{#1}\end{equation}}
\def\eeqn{\end{equation}}

\def\beqa{\begin{eqnarray}}
\def\eeqa#1{\label{#1}\end{eqnarray}}
\def\eeqan{\end{eqnarray}}

\let\bar=\overbar

\def\Dslash{\not{\hbox{\kern-4pt $D$}}}
\def\dslash{\not{\hbox{\kern-2pt $\del$}}}

\def\msb{{\bar{\ssstyle M \kern -1pt S}}}

\begin{document}
\begin{titlepage}
\pubblock

\vfill
\Title{Developing Pre-Supernova Neutrino Support for sntools}
\vfill
\Author{ Ellie O'Brien\support, Susan Cartwright, Patrick Stowell}
\Address{\sheff}
\vfill
\begin{Abstract}

The first detection of supernova burst neutrinos was achieved through the observation of SN1987A, almost four decades ago. However, neutrinos produced during the burning stages of a star prior to core collapse are yet to be detected. 
Detection of pre-supernova neutrinos could provide an early warning of an imminent supernova and allow the scientific community time to focus their resources on the observation and study of such an event leading to better understanding of these rare phenomena. Integrating pre-supernova models into a neutrino event generator would help 
to provide a unified framework for studying these neutrinos in current and 
next generation detectors. \textit{sntools} is a neutrino event generator for supernova burst neutrinos, originally developed to study supernova model discrimination with Hyper-Kamiokande. Work to add support for pre-supernova event generation to \textit{sntools} is presented, detailing the adaptations and additions to the code, with emphasis on how time binning can be optimised for a robust simulation, and also detailing the status of the validation process. The current status and capabilities of the package will be explained alongside plans for any further work and the intended use for the new functionality within the Hyper-Kamiokande Collaboration.

\end{Abstract}
\vfill
\begin{Presented}
NuPhys2026, Prospects in Neutrino Physics\\
King's College, London, UK,\\ January 7--9, 2026
\end{Presented}
\vfill
\end{titlepage}
\def\thefootnote{\fnsymbol{footnote}}
\setcounter{footnote}{0}

\section{Introduction}

When a ‘massive’ star ($>$8 $M_{\odot}$) comes to the end of its life, it explodes in the form of a Core Collapse Supernova (CCSN).
In 1987, neutrinos from a CCSN were detected for the first time. 11 supernova burst neutrinos from SN1987A were detected by Kamiokande II \cite{hirata_observation_1987}, with IMB \cite{bionta_observation_1987} 
and Baksan \cite{alekseev_detection_1987} also seeing a signal. 
Supernova burst neutrinos are produced for tens of seconds after core collapse with typical energies of tens of MeV.

Since the first detection, study of supernova burst neutrinos has steadily increased, but the study of neutrinos produced before the core collapse of a massive star, namely pre-supernova neutrinos, has only become established within the last two decades. Pre-supernova neutrinos are produced throughout all burning stages of a massive star, mainly through thermal processes such as $e^{+}e^{-}$ pair annihilation \cite{kato_theoretical_2020}. Neutrinos from silicon core and shell burning phases are produced throughout the last week before core collapse and have typical energies of only a few MeV \cite{kato_theoretical_2020} \cite{Woosley_silicon_pdf}. As first predicted in 2004 \cite{odrzywolek_detection_2004}, these neutrinos may be detectable with current and next generation neutrino detectors. 

 The instances of CCSNe in our galaxy are extremely rare and difficult to predict. The detection of pre-supernova neutrinos could provide an early warning of these events, allowing astronomers and astrophysicists valuable time to prepare to focus their resources on the observation of the star as early as possible. Much is still unknown about the explosion mechanism of supernovae and the effect of metallicity on supernova properties. It is hoped that if a nearby CCSN were to occur, a dedicated alarm system would alert the community and provide a golden opportunity to advance our knowledge and understanding of these phenomena. Pre-supernova monitoring systems and alarms have already been established by the Super-Kamiokande experiment \cite{machado_pre-supernova_2022} and the KamLAND experiment \cite{asakura_kamland_2016}, as well as a joint alarm system \cite{abe_combined_2024}. 

 During a previous pre-supernova neutrino sensitivity study for Super-Kamiokande \cite{machado_pre-supernova_2022}, it was not possible to generate Inverse Beta Decay (IBD) interactions within the detector simulation software, leading to an approach in which a weighted positron energy spectrum was fed into the software and the 2.2 MeV photon events were simulated separately and subsequently matched. This approach is complex, used approximate expressions for the IBD cross section and did not handle the determination of the direction of the outgoing positron correctly. 
Integrating pre-supernova models into an existing neutrino event generator would aid ease of use, provide complete IBD interaction event generation and help to provide a unified framework for studying these neutrinos in current and 
next generation detectors.

\section{sntools}

\textit{sntools} \cite{migenda_sntools_2021} is a neutrino event generator for supernova burst neutrinos,
originally developed to study supernova model discrimination with Hyper-Kamiokande \cite{migenda_supernova_2020} \cite{abe_supernova_2021}. It provides realistic detailed event-by-event supernova neutrino event
generation including timing and directional information for multiple
interaction channels. \textit{sntools} currently supports a range of detector geometries and detection materials and is already used by several neutrino experiments. \textit{sntools} also supports several different input formats for neutrino fluxes generated from computer simulations and the use of
core-collapse supernova models implemented in SNEWPY \cite{baxter_snewpy_2021} is enabled. 
The output file produced can be directly fed into detector simulation software to simulate the detector response to these interactions, enabling more in-depth and advanced studies than what is currently available from other related open source software packages.

\begin{figure} 
\centering
\includegraphics[width=0.9\textwidth]{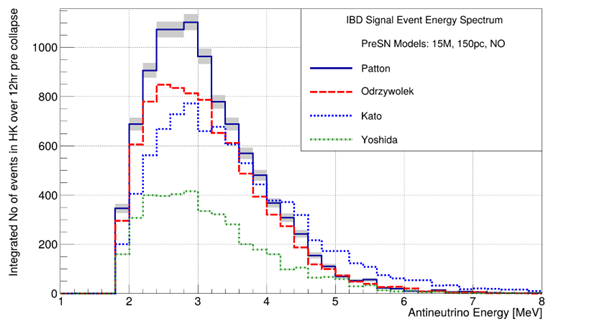}
\captionsetup{font=scriptsize, width=0.9\textwidth}
\caption{A graph of the number of IBD events in Hyper-Kamiokande generated by \textit{sntools} as a function of 
antineutrino energy for 4 different pre-supernova neutrino flux models: Odrzywolek \cite{odrzywolek_neutrino_2010}, Patton \cite{patton_presupernova_2017}, Kato \cite{kato_neutrino_2017} and Yoshida \cite{yoshida_presupernova_2016}. A standard Betelgeuse-like star is chosen as an example and normal neutrino mass ordering is assumed. Statistical error bars are shown for the Patton model. }
\label{fig:four_model_hist}
\end{figure}

\section{Pre-Supernova Neutrino Modelling}

To generate neutrinos from a CCSN neutrino burst, \textit{sntools} uses an input flux and energy spectrum to generate events within specified time bins.  As the duration of a neutrino burst is $\sim$10s, a default time bin size of 1ms was chosen to provide sufficient granularity.  This is not appropriate for pre-SN neutrinos, where the flux is lower and the time period of interest extends up to several days before core collapse.  Using 1ms bins would be inefficient, producing large numbers of empty bins and increasing simulation time unnecessarily. On the other hand, the neutrino flux increases exponentially just prior to core collapse, and using too wide a bin would not model this accurately, as \textit{sntools} calculates the neutrino flux at the bin centre. Extensive time bin optimisation studies were carried out to ensure accurate modelling of the pre-collapse flux increase while maintaining good efficiency. This yielded an optimum bin size of 1s; however, to allow flexibility, this was implemented as a default which can be overridden by the user (enabling small experiments only sensitive to the pre-collapse peak to study this in more detail). Variable bin sizes are being considered for future versions of the package. 

Following modifications to the package, the software supports four families of pre-supernova models \cite{odrzywolek_neutrino_2010} \cite{patton_presupernova_2017} \cite{kato_neutrino_2017} \cite{yoshida_presupernova_2016}, re-using their existing and well-tested implementations within SNEWPY. Figure \ref{fig:four_model_hist} shows the energy spectra of four models for a standard Betelgeuse-like star. In theory, given the significant differences between the spectra, it is possible that detection of these neutrinos by Hyper-Kamiokande could place strong constraints on the models. It is therefore essential that we have a pre-supernova neutrino event generator ready to assess sensitivity.

\section{Validation}

\begin{figure}
\centering
\includegraphics[width=0.9\textwidth]{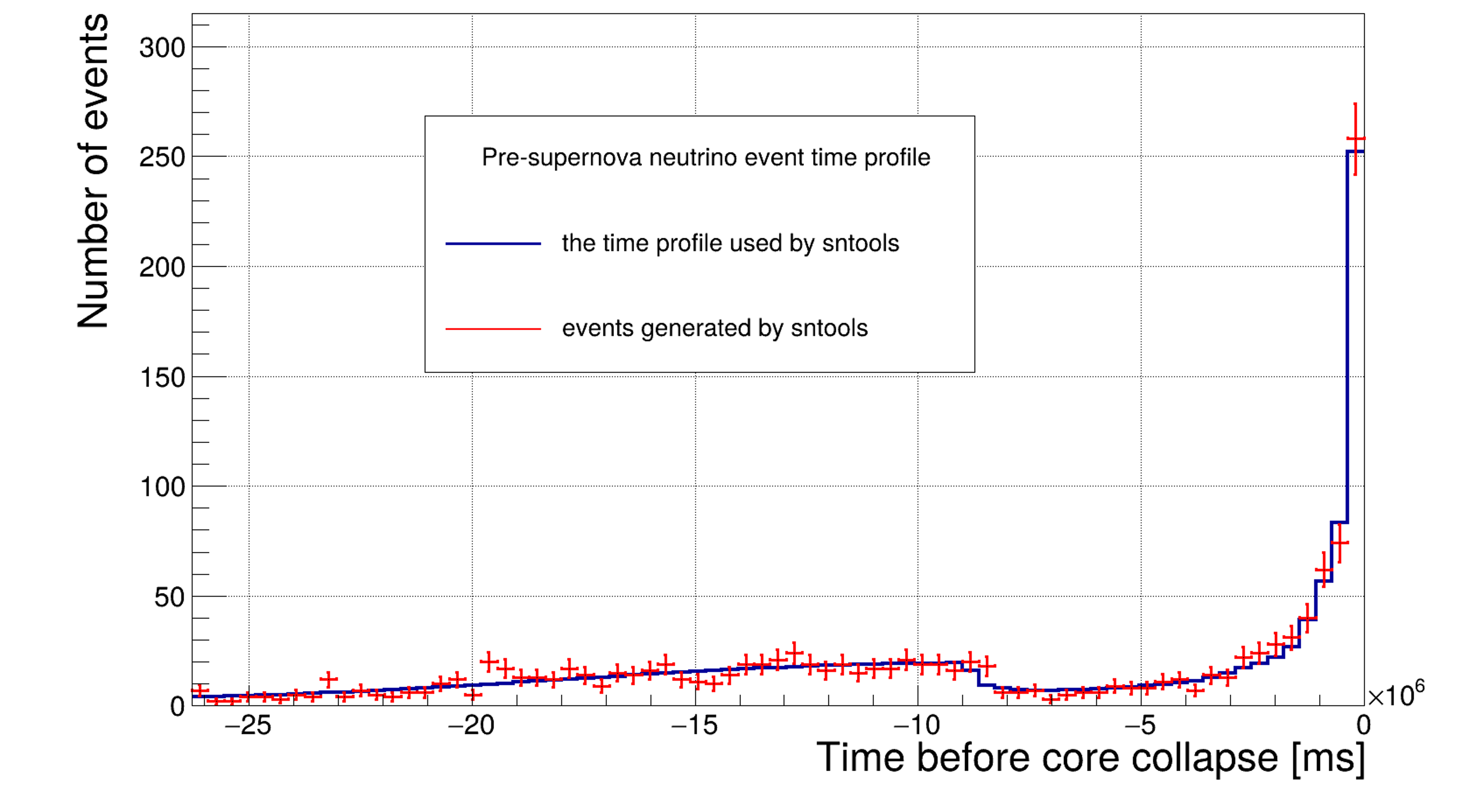}
\captionsetup{font=scriptsize, width=0.9\textwidth}
\caption{A graph showing the time profile of Inverse Beta Decay (IBD) events in the Hyper-Kamiokande geometry generated 
by \textit{sntools} compared to the expected distribution (given as an input to \textit{sntools}). The step shown approximately 2-3 hours before core collapse is caused by neutrino emission from silicon shell burning in the star.}
\label{fig:time_profile}
\end{figure}

\begin{figure}
\centering
\includegraphics[width=0.9\textwidth]{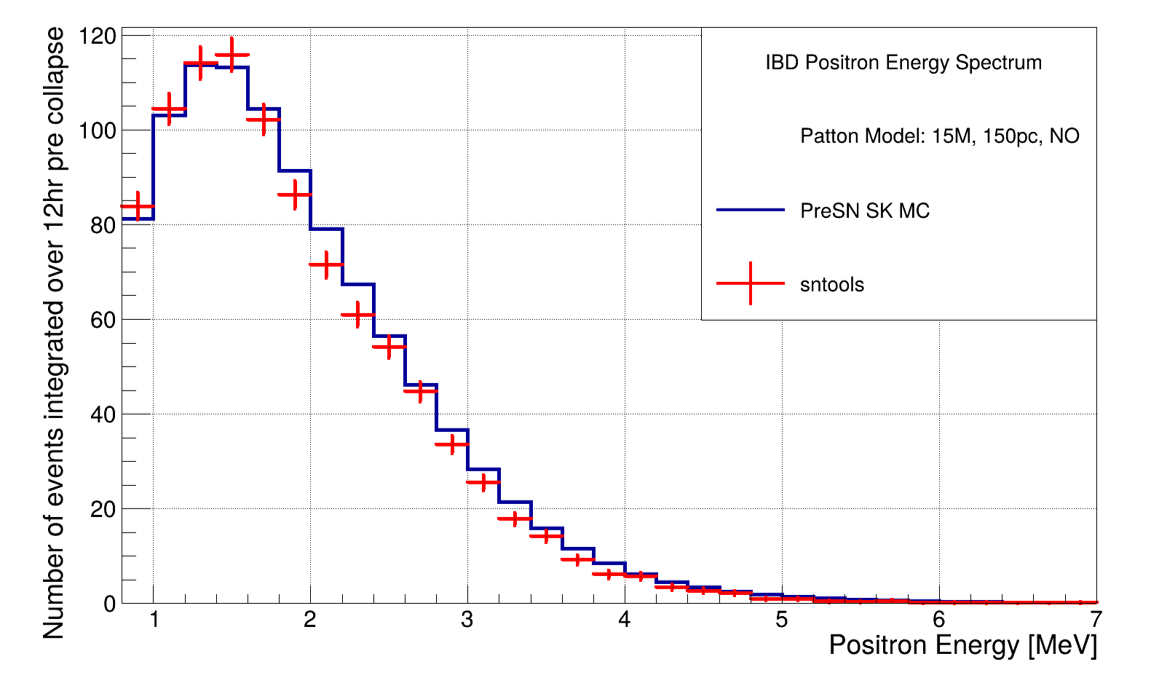}
\captionsetup{font=scriptsize, width=0.9\textwidth}
\caption{A graph showing the comparison of the positron energy spectra for IBD events 
generated by \textit{sntools} and the Super-Kamiokande pre-supernova MC events. }
\label{fig:pos_energy_hist}
\end{figure}

Before a new release of the software can be made, the properties of the events generated by \textit{sntools} are compared to those from a pre-existing method to complete a validation process. Figure \ref{fig:time_profile} demonstrates the internal consistency of \textit{sntools} and Figure \ref{fig:pos_energy_hist} shows
good agreement with Monte Carlo used for a pre-supernova
neutrino sensitivity study by Super-Kamiokande in 2022 \cite{machado_pre-supernova_2022}. Further investigation concluded that slight differences in the energy spectra featured in Figure \ref{fig:pos_energy_hist} are due to differing implementations of the (IBD) cross section and rejection sampling methods used to determine the properties of single events. Further comparisons of these methods over an expanded range of pre-supernova models, time periods and neutrino mass ordering cases is needed and the validation process is ongoing.

\section{Current Status and Further Work}

A beta version of the software (v1.5b1), including the adaptations needed to use pre-supernova models, was released in October 2025 for testing and validation purposes only. Access to the current version of \textit{sntools} is given in \cite{noauthor_snews2sntools_2025}. The completion of the validation process will allow the use of \textit{sntools} within the software chain to generate the signal Monte Carlo for a pre-supernova neutrino sensitivity study for Hyper-Kamiokande.

\Acknowledgements

We are grateful to Jost Migenda and Lucas N. Machado for their support during the development of the modifications to \textit{sntools}, the validation process and for their feedback on this work.




\printbibliography

\end{document}